\documentclass [12pt] {report}
\usepackage{amssymb}
\newcommand {\eqdef} {\stackrel{\rm def}{=}}
\newcommand {\D}[2] {\displaystyle\frac{\partial{#1}}{\partial{#2}}}
\newcommand {\DD}[2] {\displaystyle\frac{\partial^2{#1}}{\partial{#2^2}}}

\newcommand {\Dd}[3] {\displaystyle\frac{\partial^2{#1}}{\partial{#2}\partial{#3}}}

\newcommand {\Si} {\Sigma}
\newcommand {\de} {\delta}
\newcommand {\De} {\Delta}
\newcommand {\eps} {\epsilon}
\newcommand {\prtl} {\partial}
\newcommand {\fr} {\displaystyle\frac}

\newcommand {\be} {\begin{equation}}
\newcommand {\ee} {\end{equation}}
\newcommand {\ba} {\begin{array}}
\newcommand {\ea} {\end{array}}
\newcommand {\bp} {\begin{picture}}
\newcommand {\ep} {\end{picture}}
\newcommand {\bc} {\begin{center}}
\newcommand {\ec} {\end{center}}
\newcommand {\bt} {\begin{tabular}}
\newcommand {\et} {\end{tabular}}
\newcommand {\lf} {\left}
\newcommand {\rg} {\right}

\newcommand {\ses} {\medskip}
\newcommand {\pgbrk} {\pagebreak}

\newcommand {\bR} {{\bf R}}

\newcommand {\g}  {\stackrel{g\to -g}{\Longleftrightarrow}}

\newcommand {\suml} {\sum\limits}
\newcommand {\bibit} {\bibitem}
\newcommand {\nin} {\noindent}

\def\2#1#2#3{{#1}_{#2}\hspace{0pt}^{#3}}
\def\3#1#2#3#4{{#1}_{#2}\hspace{0pt}^{#3}\hspace{0pt}_{#4}}

\newcounter{sctn}
\def\sec#1.#2\par{\setcounter{sctn}{#1}\setcounter{equation}{0}
                  \noindent{\bf\boldmath#1.#2}\bigskip\par}
\begin {document}

\begin {titlepage}

\vspace{0.1in}

\begin{center}
{\Large
Finslerian Anisotropic Relativistic Metric Function Obtainable
}\\
\end{center}

\begin{center}
{\Large
under Breakdown of Rotational Symmetry
}\\
\end{center}

\vspace{0.3in}

\begin{center}

\vspace{.15in}
{\large G.S. Asanov\\}
\vspace{.25in}
{\it Division of Theoretical Physics, Moscow State University\\
117234 Moscow, Russia\\

 asanov@newmail.ru

}
\vspace{.05in}

\end{center}

\begin{abstract}

We undertake
to show how the relativistic Finslerian Metric Function (FMF)
should arise under uni-directional
violation of
spatial isotropy, keeping the condition that the indicatrix
(mass-shell) is a space of constant negative curvature.
By evaluating
respective Finslerian tetrads, and treating them consistently
as the bases of inertial reference frame
(RF), the  generalized Finslerian kinematic transformations
follow in a convenient explicit form.
The concomitant Finslerian
relativistic relations
 generalize their Lorentzian prototypes through the presence of
one characteristic parameter~$g$, so
that the constraints on the parameter may be found
in future high-precision post-Lorentzian experiments.
As the associated Finslerian Hamiltonian function is also obtainable in
a clear explicit form, convenient prospects for the Finslerian extension
of particle dynamics
are also opened.
Additionally, the Finslerian extension of the general-relativistic
Schwarzschild metric can unambiguously be proposed.
\ses

\end{abstract}

\end{titlepage}

\vskip 1cm

{\nin\bf 1. Introduction}
\medskip

Minkowski [1,2] is well-known to have
noted
that the Special Relativity (SR) implies introducing the pseudo-Euclidean
metric to geometrize the space-time. After that, the concepts of
the
SR and of
the space-time pseudo-Euclidean geometry became living as
two faces of the same cone.
This event of history impels one
to draw various farther-reaching
conclusions and, first of all,
that a researcher cannot be so self-determinent
as to believe that one is able to create new self-consistent and
universal-meaning theory
of SR without inscribing it
into the realms of some new
particular geometry introduced over
space-time.
For
a fundamental
geometry presupposition is required.
In this connection,
the
Finsler geometry
(FG) seems to
serve as being just the nearest
appropriate
metric extension of the
pseudo-Euclidean geometry.

The spatial symmetry lies
 at the root of the ordinary SR,
matching
in fact the relevant symmetry of the
pseudo-Euclidean
metric.
However,
since
beyond square-root metric one may apply ingenious methods of FG [3,4].
the Finslerian approach may be hoped to propose the post-Lorentzian
kinematic transformations compatible with the possibility of breakdown of the
spatial symmetry.
The FMF  $F$ obtained and considered below
in Section 2
refers
 to situation in which there is
an
uni-directional
breakdown.
The associated Finslerian Hamiltonian function $H$ can be derived
from $F$ in an explicit way.
The calculation of the associated curvature tensor leads to the remarkable
conclusion that the indicatrix is a space of constant negative curvature.

In Section 3 we reveal various useful relations which get simplify the
calculations involved.
Relevant co-treatment is presented in Section 4.
The meaning of the reference frame (FR) is not the same as
the meaning
of the reference
system (RS). Elucidating the distinctions among them needs the use of the
metric tensor tetrads.
They will be calculated in Section 5, which enables us to derive
in Section 6 the explicit form of the Finsler-generalized kinematic
transformations.
In the last Section 7, comments and conclusions on the general-relativistic
theoretical aspects of the problem are
presented, noting
 a handy possibility to continue Schwarzschild line element
in due Finslerian domain.
\ses
\ses

\setcounter{sctn}{2}
\setcounter{equation}{0}

{\nin\bf 2. Initial definitions and main
observations}
\ses

Suppose we are given an
$N$-dimensional vector space $V_N$. Denote by $R$ the vectors constituting
the space, so that $R\in V_N$. Any given vector $R$ assigns a particular
direction in $V_N$. Let us fix a member
$R_{(1)}$ and
 introduce the
straightline
$e_1$
oriented along the vector
$R_{(1)}$,
 and use this
$e_1$
to serve as a $R^1$-coordinate axis
in $V_N$.
In this way we get the topological product
\be
V_N=e_1\times V_{N-1}
\ee
together with the separation
\be
R=\{R^1,\bR\}, \qquad R^1\in e_1 \quad {\rm and} \quad \bR\in V_{N-1}.
\ee
Also, we introduce a pseudo-Euclidean metric
\be
q=q(\bR)
\ee
over the $(N-1)$-dimensional vector space
$V_{N-1}$.

With respect to an admissible coordinate basis $\{e_a\}$ in
$V_{N-1}$,
we obtain the coordinate representations
\be
R=\{R^p\}=\{R^0,R^a\}
\ee
and
\be
q(\bR)=\sqrt{|r_{ab}R^aR^b|},
\ee
where $r_{ab}$ are the components of a symmetric $(+-\cdots -)$-indefinite
tensor
defined over $V_{N-1}$.
The indices $(a,b,\dots)$ and
$(p,q,\dots)$ will be specified over the
ranges $(0,2,\dots,N-1)$ and $(0,1,\dots,N-1)$, respectively;
vector indices are up, co-vector indices are down; repeated up--down
indices are automatically summed; the notation $\de^a_b$ will
stand for the Kronecker symbol; $N=4$ in the physical space-time
context proper.
Sometimes we shall mention the associated post-Euclidean
metric tensor
\be
R_{pq}=\{R_{11}=-1,~R_{1a}=0,~R_{ab}=r_{ab}\}
\ee
meaningful over the whole base space $V_N$.

The approach can be converted into the dual co-framework
\be
\hat V_N=e^1\times \hat V_{N-1}
\ee
to use the separation
\be
\hat R=\{R_1,\hat{\bR}\},
\qquad
R_1\in e^1 \quad {\rm and} \quad \hat{\bR}\in \hat V_{N-1}.
\ee
With respect to the co-basis $\{e_a\}$ in
$\hat V_{N-1}$,
we obtain the coordinate representation
\be
\hat R=\{R_p\}=\{R_1,R_a\}
\ee
together with
\be
\hat q({\bf\hat R})=
\sqrt{|r^{ab}R_aR_b|},
\ee
where $r^{ab}r_{ba}=\de^a_c$.
Also,
the tensor
\be
R^{pq}=\{R^{11}=-1,~R^{1a}=0,~R^{ab}=r^{ab}\}
\ee
is
meaningful over the whole vector space $\hat V_N$.

To ease calculations we introduce the convenient notation:
\be
G= g/h,
\ee
\medskip
\be
\qquad h\eqdef\sqrt{1+\fr14g^2},
\ee
\be
g_+=-\fr12g+h, \qquad g_-=-\fr12g-h,
\ee
\medskip
\be
G_+=g_+/h\equiv -\fr12G+1, \quad G_-=g_-/h\equiv -\fr12G-1,
\ee
\medskip
\be
g^+=1/g_+=-g_-,  \qquad  g^-=1/g_-=-g_+,
\ee
\medskip
\be
g^+=\fr12g+h, \qquad g^-=\fr12g-h,
\ee
\medskip
\be
G^+=g^+/h\equiv \fr12G+1, \quad G^-=g^-/h\equiv \fr12G-1,
\ee
entailing
\be
g_++g_-=-g, \qquad g_+-g_-=2h,
\ee
\medskip
\be
g^++g^-=g, \qquad g^+-g^-=2h,
\ee
\medskip
\be
g_+g_-=-1, \qquad
g^+g^-=-1,
\ee
and
\be
(g_+)^2-(g_-)^2=
-2gh.
\ee
We have
\be
g_+\g -g_-, \qquad g^+\g -g^-, \qquad G_+\g -G_-, \qquad G^+\g -G^-.
\ee

Now, by the help of the quadratic forms
\be
B(g;R)=-\lf(R^1-g_-q\rg)\lf(R^1-g_+q\rg)
\equiv
q^2
-gqR^1
-(R^1)^2
\ee
and
\be
\hat B(g;P)=-\lf(P_1-g_+\hat q\rg)\lf(P_1-g_-\hat q\rg)
\equiv
{\hat q}^2
-g\hat qP_1
-(P_1)^2,
\ee
where
$q=q(R)$ and $\hat q=\hat q(P)$,
we introduce the FMF
\be
F(g;R)=\sqrt{|B(g;R)|}\,j(g;R),
\ee
where
\be
j(g;R)=\lf|\fr{R^1-g_-q}{R^1-g_+q}\rg|^{-G/4},
\ee
and the associated Finslerian Hamiltonian function
\be
H(g;P)=\sqrt{|\hat B(g;P)|}\,\hat j(g;P),
\ee
where
\be
\hat j(g;P)=\lf|\fr{P_1-g_+\hat q}{P_1-g_-\hat q}\rg|^{G/4}.
\ee
The functions can also be rewritten as
\be
F(g;R)=\lf|R^1-g_-q\rg|^{G_+/2}\lf|R^1-g_+q\rg|^{-G_-/2}
\ee
and
\be
H(g;P)=\lf|P_1-g_+\hat q\rg|^{G^+/2}
\lf|P_1-g_-\hat q\rg|^{-G^-/2}.
\ee

From the FMF we may calculate the associated co-vector
\be
R_p\eqdef
F(g;R)\D{F(g;R)}{R^p}
=
\fr12\D{F^2(g;R)}{R^p}
\ee
and, then, the Finslerian metric tensor
$\{g_{pq}\}$
according to
\be
g_{pq}
\eqdef \D{R_q}{R^p}.
\ee
By inserting  (2.32) in (2.33) one obtains
\be
g_{pq}(g;R)=\fr12\,\fr{\prtl^2F^2(g;R)}{\prtl R^p\prtl R^q}.
\ee
The tensor
 $\{h_{pq}\}$  given by the components
\be
h_{pq}=g_{pq}-R_pR_qF^{-2}.
\ee
is called \it the angular metric tensor\rm.
The entailed identity
\be
R^ph_{pq}=0
\ee
is useful to apply in many cases.
The associated \it Cartan torsion tensor\rm
\be
C_{pqr}(g;R)\eqdef\fr14\,\fr{\prtl^3F^2(g;R)}{\prtl R^p\prtl R^q\prtl R^r}
\ee
can also be obtained as
\rm
\be
C_{pqr}(g;R)=\fr12\,\D{g_{pq}(g;R)}{R^r}.
\ee
If we define
\be
C_p=C_p{}^q{}_r=C_{psr}g^{sr},
\ee
we find that
\be
C_p=\D{\ln\sqrt{|\det(g_{rs})|}}{R^p}.
\ee
The associated Cartan curvature tensor is
\be
S_{pqrs}=(\3Cpts\3Ctqr-\3Cptr\3Ctqs)F^2.
\ee

The Finslerian relations
appeared in this way
reduce to the pseudo-Euclidean ones in the limit
$g\to 0$; in particular,
\be
B|_{_{g=0}}=
R_{pq}R^pR^q
,
\qquad \hat B|_{_{g=0}}=
R^{pq}P_pP_q,
\ee
\bigskip
$$
j|_{_{g=0}}=1, \qquad \hat j|_{_{g=0}}= 1,
$$
\bigskip
\be
F|_{_{g=0}}= \sqrt{
|R_{pq}R^pR^q|
},
\qquad H|_{_{g=0}}= \sqrt{
|
R^{pq}P_pP_q
|
},
\ee
\bigskip
\be
g_{pq}
|_{_{g=0}}
= R_{pq},
\qquad g^{pq}|_{_{g=0}}= R^{pq},
\ee
\bigskip
\be
C_{pqr}
|_{_{g=0}}
=0.
\ee

In the timelike sector,
\be
R^0>0,\quad g_-q<R^1<g_+q,
\ee
we have
\ses

\be
B>0
\ee
and
\ses

\be
F(g;R)=\lf(R^1-g_-q\rg)^{G_+/2}\lf(g_+q-R^1\rg)^{-G_-/2},
\ee
\ses
and simple direct calculations yield
the explicit components
\ses
\ses
\be
R_1=-(R^1+gq)\fr{F^2}B,
\qquad
R_a=r_{ab}R^b\fr{F^2}B,
\ee
\ses
\ses
\be
h_{11}=-q^2\fr{F^2}{B^2},
\qquad
h_{1a}=
r_{ab}R^bR^1\fr{F^2}{B^2},
\ee
\ses
\ses
\be
h_{ab}=r_{ab}\fr{F^2}{B}
-(q-gR^1)r_{ac}R^c
r_{bd}R^d\fr{F^2}{qB^2},
\ee
together with
\ses
\be
g_{11}=[(R^1+gq)^2-q^2]\fr{F^2}{B^2},
\qquad
g_{1a}=-
gqr_{ab}R^b\fr{F^2}{B^2},
\ee
\ses
\ses
\be
g_{ab}=r_{ab}\fr{F^2}{B}
+g   r_{ac}R^c
r_{bd}R^dR^1\fr{F^2}{qB^2}.
\ee

For the reciprocal components $\{g^{pq}\}$ we find
\be
g^{11}(g;R)=-(gqR^1+B)F^{-2},
\qquad
g^{1a}(g;R)=-gqR^aF^{-2},
\ee
and
\ses
\be
g^{ab}(g;R)=BF^{-2}r^{ab}-g(R^1+gq)q^{-1}R^aR^bF^{-2},
\ee
so that $g^{pq}g_{qs}=\de^p_s$.

It follows that
\be
\det(g_{pq})=-\lf(\fr{F^2}{B}\rg)^N
\det(r_{ab}).
\ee

A convenient way to derive the Finslerian Hamiltonian function
(2.31) from the FMF (2.30) is to use the relations
\be
\hat q=q\fr{F^2}B, \qquad
R^1=-(R_1+g\hat q)\fr B{F^2},
\ee
\ses
\be
F=H,
\qquad B\hat B=F^4=F^2H^2,
\ee
and
\be
R^1=-(R_1+g\hat q)\fr{H^2}{\hat B},
\qquad
R^a=r^{ab}R_b\fr{H^2}{\hat B}.
\ee

Calculating yields
\be
F^2C_pC^p=-\lf(\fr{gN}2\rg)^2
\ee
\ses
\ses
\ses
\ses
\be
C_{pqr}=\fr1{N}\Bigl[
h_{pq}C_r
+
h_{pr}C_q
+
h_{qr}C_p
-
\fr1{C^sC_s}C_pC_qC_r\Bigr].
\ee

By inserting (2.61) in the tensor (2.41) describing the indicatrix curvature
one obtains
the result
\be
S_{pqrs}=S^*(h_{pr}h_{qs}-h_{pq}h_{rs})
\ee
with
\be
S^*=\fr14g^2.
\ee
Since
the curvature value $R_{Ind}$ of the indicatrix
(which is defined by the equation
$F=1$)
can be found from the formula
\be
R_{Ind}=-(1+S^*)
\ee
(cf. Eq. (2.3.37) in [4];
the negative sign of the signature has been taken into account),
we arrive at
\ses

\nin
{\bf PROPOSITION}. \it For the FMF \rm(2.30) \it the indicatrix
is a space of the
constant negative curvature
\be
R_{\rm Ind}=-\lf(1+\fr14g^2\rg)\le-1.
\ee
\ses
\ses

\setcounter{sctn}{3}
\setcounter{equation}{0}

{\nin\bf 3. ADDITIONAL  IMPLICATIONS}
\ses

With respect to {\it an orthogonal basis}, so that
$r_{00}=-r_{22}=\cdots =-r_{(N-1)(N-1)}=1$
(other components of the tensor
$r_{ab}$ vanish),
and in terms of the convenient notation
\be
v=R^1/R^0\equiv v^1,\quad
u=\sqrt{(R^2)^2+\cdots+(R^{N-1})^2}\,/R^0
\equiv v^{\perp},
\ee
the timelike sector
(2.46) can be specified as
\be
R^0>0,\quad
g_-M<v<g_+M,
\ee
where
\be
M=\sqrt{1-u^2}.
\ee
We introduce the functions
\be
Q(g;v,u)
=1-v^2-u^2-gvM
\equiv M^2-gvM-v^2>0
\ee
and
\ses

\be
V(g;v,u)
=\lf(v-g_-M\rg)^{G_+/2}\lf(g_+M-v\rg)^{-G_-/2}
\ee
to have
\be
B=(R^0)^2Q
\ee
and
\be
F=R^0V.
\ee

We find
\ses

\be
\D{V}{v}=-(gM+v)\fr VQ,
\qquad
\D{V}{u}=-u\fr VQ,
\ee
\ses
\ses
\ses
\ses
\be
\DD{V}{v}=-M^2\fr V{Q^2},
\qquad
\Dd{V}{v}{u}=-vu\fr V{Q^2},
\qquad
\DD{V}{u}=-\lf(1-v^2-\fr{gv}M\rg)\fr V{Q^2},
\ee
\ses
and also,
\ses

\be
\D{(V/Q)}{v}=v\fr V{Q^2},
\qquad
\D{(V/Q)}{u}=u(1-\fr{gv}{M})\fr V{Q^2},
\ee
\ses
\ses
\ses
\ses
\be
\DD{(V/Q)}{v}=(1+4v^2-u^2+gvM)\fr V{Q^3},
\ee
\ses
\ses
\ses
\be
\Dd{(V/Q)}{v}{u}=\fr uM(4vM-2gv^2+uM-gvu)\fr V{Q^3},
\ee
\ses
\ses
\ses
\be
\DD{(V/Q)}{u}=\Bigl
[
\lf(1-\fr{gv}M+\fr{gvu^2}{M^3}\rg)Q+
u^2\lf(1-\fr{gv}M\rg)\lf(3-\fr{2gv}M\rg)\Bigr]
\fr V{Q^3}.
\ee
\ses
\ses

For the function
\ses

\be
j\eqdef \fr V{\sqrt {Q}}
\ee
we get
\ses
\be
\D{j}{v}=
-\fr12gMj\fr 1Q,
\qquad
\D{j}{u}=
-\fr12g\fr{vu}Mj\fr 1Q.
\ee

The approximation of $V$ reads
\ses
\be
V\approx 1-gv-\fr12(v^2+u^2)-\fr16gv^3-\fr18(v^2+u^2)^2-\fr1{12}g^2v^4+...\,,
\ee
so that
\ses
\be
\fr VQ\approx 1+\fr12(v^2+u^2)+\fr16gv^3-gvu^2+... \, .
\ee
\ses
\ses

\setcounter{sctn}{4}
\setcounter{equation}{0}

{\nin\bf 4. CO-FRAMEWORK}
\ses

We similarly proceed for the co-vectors in the time-like sector:
\be
\hat v
=
-\fr{P_1}{P_0}
=\hat v^1
,\quad
\hat u=\sqrt{(P_2)^2+\cdots +(P_{N-1})^2}\,/P_0
\equiv \hat v^{\perp},
\ee
\ses
\be
P_0>0,
\qquad
-g_+\hat M<\hat v<-g_-\hat M,
\ee
\ses
\be
\hat Q
(g;\hat v,\hat u)
\eqdef\hat B/(P_0)^2,
\qquad
\hat M\eqdef\hat q/P_0.
\ee
We have
\be
\hat M
=
\sqrt{1-{\hat u}^2},
\ee
\ses
\be
W\eqdef H/P_0,
\ee
\ses
\be
\hat Q
(g;\hat v,\hat u)
=1-{\hat v}^2-{\hat u}^2+g\hat v\hat M
\equiv {\hat M}^2+g\hat v\hat M-{\hat v}^2>0,
\ee
\ses
\be
\hat Q(g;\hat v,\hat u)
=(g_+\hat M+\hat v)(-g_-\hat M-\hat v),
\ee
and
\ses
\be
W(g;\hat v,\hat u)
=\lf(      g_+\hat M+\hat v\rg)^{-G_-/2}\lf(-g_-\hat M-\hat v\rg)^{G_+/2}.
\ee
\ses\\
By differentiating (4.8)
we find
\ses

\be
\D{W}{\hat v}=(-\hat v+g\hat M)\fr W{\hat Q},
\qquad
\D{W}{\hat u}=-\hat u\fr W{\hat Q},
\ee
Also,
\ses

\be
\D{(W/\hat Q)}{\hat v}=\hat v\fr W{{\hat Q}^2},
\qquad
\D{( W/\hat Q)}{\hat u}=\hat u(1+\fr{g\hat v}{\hat M})\fr W{{\hat Q}^2}.
\ee

For the function
\ses

\be
\hat j\eqdef \fr W{\sqrt {\hat Q}}
\ee
we get
\ses

\be
\D{\hat j}{\hat v}=
\fr12g\hat M\hat j\fr 1{\hat Q},
\qquad
\D{\hat j}{\hat u}=
\fr12g\fr{\hat v\hat u}{\hat M}\hat j\fr 1{\hat Q}.
\ee

The approximation of $W$ reads
\ses
\be
W\approx 1+g\hat v-\fr12({\hat v}^2+{\hat u}^2)+\fr16g{\hat v}^3
-\fr18({\hat v}^2+{\hat u}^2)^2-\fr1{12}g^2{\hat v}^4+...\, .
\ee
\ses

\be
\fr W{\hat Q}\approx 1+\fr12({\hat v}^2+{\hat u}^2)
-\fr16g\hat v^3+g\hat v\hat u^2+...
\ee
These differ formally
from (3.16)-(3.17) in the negative sign placed at the parameter  $g$.

Considering the dispersion relation
\be
P_0W=m,
\ee
we find
\ses
\ses
\be
\D{P_0}{P_1}=
\fr{P_1+g
\sqrt
{(P_0)^2-(P_{\perp})^2}
}
{
P_0
}
\ee
and
\ses
\be
\D{P_0}{P_{\perp}}=
\fr{|P_{\perp}|}
{
P_0}.
\ee
So we get
the approximation
\be
P_0\approx m+gP_1+\fr1{2m}
((P_0)^2+(P_{\perp})^2)+...\,.
\ee
\ses
\ses

\setcounter{sctn}{5}
\setcounter{equation}{0}
\ses

{\bf\nin 5. POST-LORENTZIAN  TETRADS}
\ses

The associated
Finslerian
metric tensor $g_{pq}(g;R)$ given by the components (2.52) and (2.53)
proves to be of space-time
 signature, that is, be representable as follows:
\be
g_{pq}(g;R)=
\suml_{P=0}^{N-1}q_{(P)}H_p^{(P)}(g;R)H_q^{(P)}(g;R),~~q_{(P)}=(1,-1,\cdots,-1).
\ee
The contravariant tetrad reciprocal to $H^{(P)}_p(g;R)$ will be denoted
by $H^p_{(P)}(g;R)$, so that $H^{(P)}_p(g;R)H^p_{(Q)}(g;R) =\de_Q^P$.
The Finslerian
tetrad $H^{(P)}_p(g;R)$ is a geometrical
representation
for a RS of a RF.
If
\be
H_{(0)}^p(g;R)||R^p,
\ee
we call the tetrad \it  proper kinematical\rm, for
in such a case
the vector $H_{(0)}^p$ indicates the four-dimensional direction
along which the local observer is moving.

On using (2.52)-(2.55) and (3.1)-(3.6),
the attentative
calculation of tetrads obeying (5.1)-(5.2) results in the following
list:
\be
H_{(0)}^0(g;v,u)=\fr{1}{V(g;v,u)},
\quad
H_{(0)}^1(g;v,u)=\fr{v}{V(g;v,u)},
\quad
H_{(0)}^2(g;v,u)=\fr{u}{V(g;v,u)},
\ee
\ses
\ses
\be
H_{(1)}^0(g;v,u)=
(\fr{v}M+g)
\fr{1}{V(g;v,u)},
\quad
H_{(1)}^1(g;v,u)=
\fr{M}{V(g;v,u)},
\quad
H_{(1)}^2(g;v,u)=
(\fr{v}M+g)
\fr{u}{V(g;v,u)},
\ee
\ses
\ses
\be
H_{(2)}^0(g;v,u)=
\fr{u}{M}\fr{\sqrt{Q(g;v,u)}}{V(g;v,u)},
\quad
H_{(2)}^1(g;v,u)=0,
\quad
H_{(2)}^2(g;v,u)=
\fr{1}{M}\fr{\sqrt{Q(g;v,u)}}{V(g;v,u)},
\ee
\ses
\ses
\be
H_{(3)}^0(g;v,u)=H_{(3)}^1(g;v,u)=H_{(3)}^2(g;v,u)=0,
\quad
H_{(3)}^3(g;v,u)=
\fr{\sqrt{Q(g;v,u)}}{V(g;v,u)},
\ee
\ses
\ses
\ses
\be
H_{(0)}^3(g;v,u)=H_{(1)}^3(g;v,u)=H_{(2)}^3(g;v,u)=0,
\ee
\ses
together with
\be
H^{(0)}_0(g;v,u)=
\fr{V(g;v,u)}
{Q(g;v,u)},
\quad
H^{(0)}_1(g;v,u)=-(v+gM)
\fr
{V(g;v,u)}
{Q(g;v,u)},
\quad
H^{(0)}_2(g;v,u)=-u
\fr
{V(g;v,u)}
{Q(g;v,u)},
\ee
\ses
\ses
\be
H^{(1)}_0(g;v,u)=
-\fr{v}{M}
\fr
{V(g;v,u)}
{Q(g;v,u)},
\quad
H^{(1)}_1(g;v,u)=M
\fr
{V(g;v,u)}
{Q(g;v,u)},
\quad
H^{(1)}_2(g;v,u)=
\fr{vu}{M}
\fr
{V(g;v,u)}
{Q(g;v,u)},
\ee
\ses
\ses
\be
H^{(2)}_0(g;v,u)=
-
\fr{u}{M}
\fr
{V(g;v,u)}
{\sqrt{Q(g;v,u)}},
\quad
H^{(2)}_1(g;v,u)=
0,
\quad
H^{(2)}_2(g;v,u)=
\fr{1}{M}
\fr
{V(g;v,u)}
{\sqrt{Q(g;v,u)}},
\ee
\ses
\ses
\be
H^{(3)}_0(g;v,u)=
H^{(3)}_1(g;v,u)=
H^{(3)}_2(g;v,u)=
0,
\quad
H^{(3)}_3(g;v,u)=
\fr
{V(g;v,u)}
{\sqrt{Q(g;v,u)}},
\ee
\ses
\ses
\ses
\ses
\ses
\be
H^{(0)}_3(g;v,u)=H^{(1)}_3(g;v,u)=H^{(2)}_3(g;v,u)=0.
\ee
\ses
\ses

\setcounter{sctn}{6}
\setcounter{equation}{0}
\ses

{\bf\nin 6. EXPLICATED FINSLERIAN  KINEMATIC  TRANSFORMATIONS}
\ses

\nin
If $A^p$ denotes a four-dimensional vector
that represents the motion
velocity of
the RF
$S_{\{v,u\}}$
with respect to
the RF
$
\Si
$, so that
\be
v\equiv v^1=\fr{\De A^1}{\De A^0}, \qquad u\equiv v^2= \fr{\De A^2}{\De A^0},
\qquad v^3=0
\ee
(when motion is going in the plane $R^1\times R^2$)
and the time-like sector is chosen, in accordance with (3.2) and (3.4),
we may use the tetrads of preceding section to connect the RFs:
\be
R^p=H_{(Q)}^p(g;A)r^{(Q)}
\ee
(the tetrads play the role of the reference systems associated with
the RFs).
In this way,
straightforward calculations 
lead to the sought generalized kinematic transformations.

Namely, on re-labeling
$r^{(Q)}=(t,x,y,z)$ and using the list (5.3)-(5.12),
with $A$ substituted with $R$, we get explicitly
\be
R^0=
\fr{1}{V(g;v,u)}
\Bigl[t+
(\fr vM+g)x+
\fr{u}{M}\sqrt{Q(g;v,u)}\,y\Bigr],
\ee
\ses
\ses
\ses
\be
R^1=
\fr{1}{V(g;v,u)}
(vt+
Mx)
,
\ee
\ses
\ses
\ses
\be
R^2=
\fr{1}{V(g;v,u)}
\Bigl[ut+
(\fr vM+g)
ux+
\fr{1}{M}\sqrt{Q(g;v,u)}\,y\Bigr],
\ee
\ses
\ses
\ses
\be
R^3
=
\fr{\sqrt{Q(g;v,u)}}{V(g;v,u)}z,
\ee
\ses

\nin
which inverse reads
\ses
\ses

\be
t=
\fr{V(g;v,u)}
{Q(g;v,u)}\Bigl[R^0-(v+gM)R^1-uR^2\Bigr],
\ee
\ses
\ses
\ses
\be
x=
\fr{V(g;v,u)}
{Q(g;v,u)}\Bigl[-\fr vMR^0+MR^1+\fr{vu}MR^2\Bigr],
\ee
\ses
\ses
\ses
\be
y=
\fr
{V(g;v,u)}
{\sqrt{Q(g;v,u)}}\fr1M(-uR^0+R^2),
\ee
\ses
\ses
\ses
\be
z=
\fr
{V(g;v,u)}
{\sqrt{Q(g;v,u)}}R^3.
\ee
\ses

\nin
Here, $Q$ and $V$ are the functions (3.4) and (3.5), respectively;
\be
\{R^0,R^1,R^2,R^3\}\in\Si, \qquad
\{t,x,y,z\}\in
S_{\{v,u\}},
\ee
where
$\Si$
is the input preferred rest frame and
$S_{\{v,u\}}$ is an inertial RF moving with the velocity
$\{v^1=v, v^2=u, v^3=0\}$ relative to
$\Si$;
the
instantaneously
common origin of the frames being implied.

Similarly for the momenta,
\ses

\be
P_0=
\fr{V(g;v,u)}
{Q(g;v,u)}
\Bigl[p_0-
\fr vMp_1-
\fr{u}{M}\sqrt{Q(g;v,u)}\,p_2\Bigr],
\ee
\ses
\ses
\ses
\be
P_1=
\fr{V(g;v,u)}
{Q(g;v,u)}
\Bigl[-(v+gM)p_0+
Mp_1
\Bigr],
\ee
\ses
\ses
\ses
\be
P_2=
\fr{V(g;v,u)}
{Q(g;v,u)}
\Bigl[-up_0+
\fr{ vu}Mp_1+
\fr{1}{M}\sqrt{Q(g;v,u)}\,p_2\Bigr],
\ee
\ses
\ses
\ses
\be
P_3=
\fr
{V(g;v,u)}
{\sqrt{Q(g;v,u)}}p_3,
\ee
and its inverse
\ses

\be
p_0=
\fr{1}{V(g;v,u)}
(P_0+
 vP_1+u
P_2),
\ee
\ses
\ses
\ses
\be
p_1=
\fr{1}{V(g;v,u)}
\Bigl[
(\fr vM+g)
P_0+
MP_1
+
(\fr vM+g)
u
P_2\Bigr],
\ee
\ses
\ses
\be
p_2=
\fr
{
\sqrt{Q(g;v,u)
}}
{V(g;v,u)}
\fr 1M
(uP_0+P_2)
\ee
\ses
\ses
\ses
\be
p_3
=
\fr{\sqrt{Q(g;v,u)}}{V(g;v,u)}P_3,
\ee
where
\ses

\be
\{P_0,P_1,P_2,P_3\}\in\Si, \qquad
\{p_0,p_1,p_2,p_3\}\in S_{\{v,u\}}.
\ee
\ses

\nin
The invariance of the contraction:
\be
R^0P_0+R^1P_1+R^2P_2+R^3P_3
=
tp_0+xp_1+yp_2+zp_3
\ee
can be verified directly.

Inversely, one may postulate (6.21) to explicate (6.12)-(6.19)
from (6.3)-(6.10).

The transformations (6.3)-(6.10) can also be rewritten in the form
\be
t=a(g;v,u)R^0+\eps(g;v,u) x+\eps_2(g;v,u)y,
\ee
\ses
\be
x=b(g;v,u)R^1+b_1(g;v,u) R^0+b_2(g;v,u)R^2,
\ee
\ses
\be
y=\fr{d(g;v,u)}{
\sqrt{1-u^2}
}(R^2-uR^0),
\ee
and
\be
z={d(g;v,u)}
R^3,
\ee
with the kinematic coefficients
\be
a=V(g;v,u),
\ee
\ses
\be
\eps=-\fr{v}{
\sqrt{1-u^2}
}
+g,
\ee
\ses
\be
\eps_2=-\fr{u}{
\sqrt{1-u^2}}\sqrt{Q(g;v,u)
}\,,
\ee
\ses
\be
b=\fr{V(g;v,u)}{Q(g;v,u)}\sqrt{1-u^2},
\quad
b_1=-\fr{V(g;v,u)}{Q(g;v,u)}\fr{v}{\sqrt{1-u^2}},
\ee
\ses
\be
b_2=\fr{V(g;v,u)}{Q(g;v,u)}\fr{vu}{\sqrt{1-u^2}}\,,
\ee
and
\ses
\be
d=\fr{V(g;v,u)}{\sqrt{Q(g;v,u)}}.
\ee

Whenever
$|v|\ll1$
and
$|u|\ll1$,
{\nin\it  the low-velocity approximations}

\be
a(g;v,u)
\approx
1-gv-\fr12(v^2+u^2)-\fr16gv^3-\fr18(v^2+u^2)^2-\fr1{12}g^2v^4+...,
\ee
\ses
\be
b(g;v,u)
\approx 1+\fr12v^2+\fr16gv^3-gvu^2+...,
\ee
\ses
\be
d(g;v,u)
\approx 1-\fr12gv+...,
\ee
\ses
\be
\eps(g;v,u)
\approx g-v-\fr12vu^2+...,
\ee
\ses
\be
\eps_2(g;v,u)
\approx -u+\fr12gvu+...,
\ee
\ses
\be
b_1(g;v,u)
\approx -v
(1+\fr12v^2+u^2+\fr16gv^3-gvu^2+...),
\ee
\ses
\be
b_2(g;v,u)
\approx vu
(1+\fr12v^2+u^2+\fr16gv^3-gvu^2+...)
\ee
are obtained.

The Finslerian parameter $g$ is characteristic, so that
in the limit $g\rightarrow 0$
the above kinematic
transformations
(6.3)-(6.10)
reduce exactly to the ordinary special-relativistic
Lorentz transformations
\be
R^0=\fr{1}{\sqrt{1-v^2-u^2}}t+\fr v{\sqrt{1-v^2-u^2}}x,
\qquad R^1=\fr v{\sqrt{1-v^2-u^2}}t+\fr{1}{\sqrt{1-v^2-u^2}}x,
\ee
\bigskip
\be
R^2=y,\qquad R^3=z,
\ee
\bigskip
\be
t=\fr1{\sqrt{1-v^2-u^2}}R^0-\fr v{\sqrt{1-v^2-u^2}}R^1,
\qquad  x=-\fr v{\sqrt{1-v^2-u^2}}R^0+\fr1{\sqrt{1-v^2-u^2}}R^1,
\ee
\ses
\ses
\ses
\ses
\ses\\
and
\bigskip
\be
P_0=\fr1{\sqrt{1-v^2-u^2}}p_0-\fr v{\sqrt{1-v^2-u^2}}p_1,
\qquad  P_1=-\fr v{\sqrt{1-v^2-u^2}}p_0+\fr1{\sqrt{1-v^2-u^2}}p_1,
\ee
\bigskip
\be
P_2=p_2,\qquad P_3=p_3,
\ee
\bigskip
\be
p_0=\fr{1}{\sqrt{1-v^2-u^2}}P_0+\fr v{\sqrt{1-v^2-u^2}}P_1,
\qquad p_1=\fr v{\sqrt{1-v^2-u^2}}P_0+\fr{1}{\sqrt{1-v^2-u^2}}P_1.
\ee
\ses
\ses

\setcounter{sctn}{7}
\setcounter{equation}{0}

{\nin\bf 7. Concluding Remarks}
\bigskip

\nin
We have served the purpose of organizing the FG-theoretical framework
which goes
beyond square-root metric and admits a single
preferred spatial direction,
thereby offering rather general theoretical and methodological basis for
anisotropic
relativistic applications.

Apart from relevant
 generalized-kinematic applications, which can be developed
with the aid of the
post-Lorentzian Finslerian kinematic transformations (6.3)-(6.10),
there are also opening up extended general-relativistic prospects,
as being suggested by the very form of the FMF (2.30) under study.
Indeed, the later function invites lifting the
Schwarzschild line element
\be
\lf(ds^{\stackrel{Schwarzschildean}
{isotropic}}\rg)^2
=
(1-2U)c^2(dt)^2-\fr1{1-2U}(dr)^2-
r^2\left((d\theta)^2+\sin^2\theta (d\phi)^2\right)
\ee
($U=KM_0/c^2r$ and
$K$ is the universal gravitational constant;
the spherical coordinates have been used;
the gravitational field is created by a mass $M_0$ at the origin)
to the advanced-Finslerian level by introducing
the extension
\be
\lf(ds^{\stackrel{Finslerian}
{anysotropic}}\rg)^2
=
\Biggl[\fr1{\sqrt{1-2U}}dr-g_-\sqrt{(1-2U)c^2(dt)^2
- r^2\left((d\theta)^2+\sin^2\theta (d\phi)^2\right)}\,
\Biggr]^{G_+}\times
\ee
\ses
\be
\Biggl[
g_+\sqrt{(1-2U)c^2(dt)^2
- r^2\left((d\theta)^2+\sin^2\theta (d\phi)^2\right)}
-\fr1{\sqrt{1-2U}}dr
\,
\Biggr]^{-G_-}
\ee
in accord with the correspondence principle
to hold true in the limit
$g\to 0$:
\be
ds^{\stackrel{Finslerian}
{anisotropic}}
|_{_{g=0}}=
ds^{\stackrel{Schwarzschildean}
{isotropic}}
\ee

Alternatively,
for the FG-isotropic way developed in the previous papers [5-9],
one gets
\be
\lf(ds^{\stackrel{Finslerian}
{isotropic}}\rg)^2
=
\Biggl[\sqrt{1-2U}cdt+g_-\sqrt{\fr1{1-2U}(dr)^2
+ r^2\left((d\theta)^2+\sin^2\theta (d\phi)^2\right)}\,
\Biggr]^{G_+}\times
\ee
\ses
\be
\Biggl[\sqrt{1-2U}cdt+g_+\sqrt{\fr1{1-2U}(dr)^2
+ r^2\left((d\theta)^2+\sin^2\theta (d\phi)^2\right)}\,
\Biggr]^{-G_-},
\ee
\ses

\nin
for which again
\be
ds^{\stackrel{Finslerian}
{isotropic}}
|_{_{g=0}}=
ds^{\stackrel{Schwarzschildean}
{isotropic}}.
\ee

All the Finslerian
relativistic relations derived can, in principle,
be verified experimentally
to put empirical limitations on the characteristic Finslerian parameter
$g$.
The corresponding applications will be made in
a special work to be published elsewhere.

Generally, as the flow of publications concerning post-Lorentzian effects
is increasing steadily,  and  the
reasons for such effects are claimed
starting from
various
interesting and important standpoints [10-32],
it might be useful to coordinate relevant attempts with the possibility of
constructing
post-Lorentzian
extensions in a FG-consistent geometrical rigorous way.

\pgbrk

\def\bibit[#1]#2\par{\rm\noindent\parskip1pt
                     \parbox[t]{.05\textwidth}{\mbox{}\hfill[#1]}\hfill
                     \parbox[t]{.925\textwidth}{\baselineskip11pt#2}\par}

\nin{\bf References}
\bigskip

\bibit[1] H. Minkowski: \it Raum und Zeit~--- Phys. Z. \bf10 \rm(1909), 104.

\bibit[2] H. Minkowski: \it Das Relativit\"atsprinzip~--- Ann. Physik. \bf47
\rm(1915), 927.

\bibit[3] H. Rund: \it The Differential Geometry of Finsler
 Spaces, \rm Springer, Berlin 1959.

\bibit[4] G.S. Asanov: \it Finsler Geometry, Relativity and Gauge
 Theories, \rm D.~Reidel Publ. Comp., Dordrecht 1985.

\bibit[5] G.S. Asanov: \it Rep. Math. Phys.
\bf 39 \rm (1997), 69;
\bf 41 \rm (1998), 117;
\bf 46 \rm (2000), 383;
\bf 47 \rm (2001), 323.

\bibit[6] G.S. Asanov: \it Moscow University Physics Bulletin
 \bf49\rm(1) (1994),~18;
\bf49\rm(2) (1994),~11;
\bf51\rm(1) (1996),~15;
\bf51\rm(2) (1996)~6;
\bf51\rm(3) (1996)~1;
\bf53\rm(1) (1998),~15.

\bibit[7] G.S. Asanov: The Finsler-type recasting of Lorentz transformations.
In: Proceedings of Conference ``Physical Interpretation of
Relativity Theory", September 15-20, London, Sunderland, 2000, pp. 1-24.

\bibit[8] G.S. Asanov: Can neutrinos and high-energy particles test
Finsler metric of space-time?
arXiv:hep-ph/0009305, 27 Sep. 2000.

\bibit[9] G.S. Asanov:
\it Found. Phys. Lett. \bf 15 \rm (2002), 199.

\bibit[10] Proceedings of Conference ``Physical Interpretation of
Relativity Theory", September 15-20, London, Sunderland, 2000.

\bibit[11] J.A. Winnie: \it Philos. Sci. \bf37 \rm(1970), 81, 223.

\bibit[12] G. Spavieri: \it Found. Phys. Lett. \bf 1 \rm (1988), 373.

\bibit[13] S.J. Prokhovnik and W.T. Morris: \it Found. Phys.
\bf 19 \rm(1989), 531.

\bibit[14] R. Mansouri and R. Sexl: \it Gen. Rel. Grav. \bf8 \rm(1977), 496, 515, 809.

\bibit[15] S.J. Prokhovnik: \it Found. Phys. \bf 3 \rm(1973), 351; \bf9
\rm(1979), 883; \bf10 \rm(1980, 197; \bf19 \rm(1989), 541; \it J.~Australian
Math. Soc. \bf5\rm(2) (1965), 273; \bf6\rm(1) (1966), 101.

\bibit[16] S.J. Prokhovnik: \it The Logic of Special Relativity, \rm
Cambridge University Press, Cambridge, 1967.

\bibit[17] A.K.A. Maciel and J. Tiomno: \it Found. Phys. \bf 19 \rm(1989), 505 and 521.

\bibit[18] G. Spavieri: \it Phys. Rev. \bf A34 \rm(1986), 1708.

\bibit[19] W.A. Rodrigues and J. Tiomno: \it Found. Phys. \bf 15 \rm(1985), 945.

\bibit[20] Y.Z. Zhang: \it Gen. Rel. Grav. \bf27 \rm(1995), 475.

\bibit[21] F. Selleri: \it Chin. J. Eng. Elect. \bf 6
\rm(1995), 25;
\it Found. Phys. Lett. \bf 9\rm(1) (1997), 73;
\it Found. Phys. \bf26 \rm(1996), 641.

\bibit[22] F. Selleri: ``Space and Time should be preferred to
Spacetime, 1 and 2". In: Redshift and gravitation in a relativistic universe
(K. Rudnicki, ed.), Apeiron, Montreal (2001), pp. 63-71 and81-94.

\bibit[23] M.D. Gabriel and M.P. Haugan: \it Phys. Rev. \bf D41 \rm(1990), 2943.

\bibit[24] M.P. Haugan and C.M. Will: \it Physics Today \bf 40 \rm(1987), 69.

\bibit[25] C.M. Will: \it Phys. Rev. \bf D45 \rm(1992), 403.

\bibit[26] S. Marinov: \it Czech. J. Phys. \bf24 \rm(1974), 965; \it Found. Phys.
\bf9 \rm(1979),445; \it Gen. Rel. Grav. \bf12 \rm(1980), 57.

\bibit[27] E.M. Kelly: \it Found. Phys. \bf14, \rm(1984), 705; \bf15 \rm(1985), 333.

\bibit[28] A. Brillet and J.L. Hall: \it Phys. Rev. Lett. \bf 42 \rm(1979), 549.

\bibit[29] S. Coleman and S.L. Glashow: Cosmic ray and neutrino tests of
special relativity
\it Harvard University Report \rm No. HUTP-97/A008, hep-ph/9703240, 1-5.

\bibit[30] S.L. Glashow, A. Halprin, P.L. Krastev, C.N. Leung, and J. Pantaleone:
Remarks on  neutrino tests of special relativity. Phys. Rev.D \bf56
\rm ,N 4 (1997), 2433-2434.

\bibit[31] R. Anderson, I. Vetharaniam and G.E. Stedman: \it  Phys. Reports
\bf 295, \rm(1998), 93-180.

\bibit[32] E. Minguzzi:
\it Found. Phys. Lett. \bf 15 \rm (2002), 153.

\end {document}